\documentclass[12pt]{article}
\usepackage[utf8]{inputenc}
\usepackage{amsfonts,amsmath}
\usepackage[left=1.5cm, right=1.5cm, top=2cm, bottom=2cm, headsep=0.7cm, footskip=1cm
]{geometry}
\newtheorem{theorem}{Theorem}

\begin{document}

\title{Two-Dimensional Finite-Gap Schr\"odinger Operators as Limits of Two-Dimensional Integrable Difference Operators\thanks{Research supported by the Russian Science Foundation grant No. 24-11-00281, https://rscf.ru/project/24-11-00281/.}}

\author{P. A. Leonchik, G. S. Mauleshova, A. E. Mironov}

\date{}
\maketitle

\begin{abstract}
In this paper we study two-dimensional discrete operators whose eigenfunctions at zero energy level are given by rational functions on spectral curves. We extend discrete operators to difference operators and show that two-dimensional finite-gap Schr\"odinger operators at fixed energy level can be obtained from difference operators by passage to the limit.

\end{abstract}

\section{Introduction}

This work continues our investigation of the relationship between integrable differential and difference operators. In \cite{MM}, we discovered the following phenomenon: one-dimensional finite-gap Schr\"odinger operator commuting with ordinary differential operator of odd order $2g+1$ can be extended to difference operator depending on a small parameter that commutes with difference operator of the same order $2g+1$. Meanwhile, all fundamental properties of the differential and difference operators coincide. Specifically: the spectral curve of differential operators coincides with the difference operators' spectral curve and is independent of the small parameter; the maximal commutative ring of differential operators containing the Schr\"odinger operator is isomorphic to the maximal commutative ring of difference operators containing the second-order difference operator (i.e. these operators are one-point \cite{KN}). This result holds more generally: ordinary commuting differential operators of rank one can be extended to one-point difference operators of rank one, with Krichever's spectral data naturally extending to difference operator spectral data, and as the small parameter tends to zero, the difference operators converge to differential ones.

In this work, we establish analogous results for two-dimensional finite-gap Schr\"odinger operators at fixed energy level

\begin{equation} 
\label{eq1}
H = \partial_z \partial_{\bar{z}} + A(z,\bar{z}) \partial_{\bar{z}} + u(z,\bar{z}),
\end{equation}

introduced by Dubrovin, Krichever, and Novikov \cite{DKN}. The kernel of $H$ contains the two-point Baker-Akhiezer function $\varphi(z,\bar{z},P)$, $H\varphi=0$, defined on a Riemann surface $\Gamma$ of genus $g$, where $P\in\Gamma$. The function $\varphi$ has two essential singularities on $\Gamma$, as well as $g$ simple poles independent of $z$ and $\bar{z}$. In \cite{NV}, a condition was established for the operator $H$ to be potential (i.e., $A = 0$), while \cite{GMN1}, \cite{GMN2} derived the condition for $H$ to be factorizable. Finite-gap Schr\"odinger operators arise in various areas of mathematical physics and geometry, for instance, in the construction of tori with constant mean curvature in  ${\mathbb R}^3$ and minimal Lagrangian tori in ${\mathbb C}P^2$. 

In \cite{LM}, a class of discrete operators of the form
	\begin{equation} 
	\label{eq2}
		L=T_1 T_2 + a_{n,m} T_1 + b_{n,m} T_2 + v_{n,m},
	\end{equation}
was obtained, where $T_1,T_2$ --- shift operators $T_1f(n,m)=f(n+1,m), \ T_2f(n,m)=f(n,m+1),$ such that the kernel of $L$ contains a discrete Baker--Akhiezer function $\psi(n,m,P)$ satisfying $L \psi(n,m,P) = 0$. For fixed $n,m$, the function $\psi$ is rational on the spectral curve $\Gamma$; in particular, like the Baker--Akhiezer function $\varphi$, it has simple poles at $g$ points independent of $n$ and $m$. Periodic operators $L$ were considered in \cite{K} (see also \cite{DGNS}).
	
In this work, we explicitly find discrete Baker--Akhiezer functions $\psi(n,m,P)$ in terms of the theta functions of the Jacobian variety of the spectral curve $\Gamma$ (Theorem 2) and extend the operator (\ref{eq2}) to a difference operator of the form
	\begin{equation} 
	\label{eq3}
L_{\varepsilon,\delta} = \frac{T_{\varepsilon}}{\varepsilon} \frac{T_{\delta}}{\delta} + a(z_1,z_2) \frac{T_{\varepsilon}}{\varepsilon} + b(z_1,z_2) \frac{T_{\delta}}{\delta} + v(z_1,z_2),
	\end{equation}
where $T_{\varepsilon}, T_{\delta}$ --- shift operators on $\varepsilon$ and $\delta$ respectively, $T_{\varepsilon}f(z_1,z_2)=f(z_1+\varepsilon,z_2),$ \ $T_{\delta}f(z_1,z_2)=f(z_1,z_2+\delta),$ \ $\varepsilon, \delta \in \mathbb{C},$ coefficients $a(z_1,z_2), \ b(z_1,z_2), \ v(z_1,z_2)$ are expressed in terms of theta functions. For $z_1 = z$, $z_2=\bar{z}$ and $\varepsilon,\delta \to 0$, the operator $L_{\varepsilon,\delta}$ converges to $H$ (Theorem 3). The restriction of operator $L_{\varepsilon,\delta}$ to the lattice ${\varepsilon \mathbb{Z}, \delta \mathbb{Z}}$ contains a discrete Baker--Akhiezer function in its kernel.

In Section 2, we recall some results from \cite{MM} concerning finite-gap one-dimensional Schr\"odinger operators and commuting difference operators, as well as results from \cite{DKN} about finite-gap two-dimensional Schr\"odinger operators at one energy level. Two-dimensional discrete operators will be considered in Section 2.3, where Theorem 2 is proved. In Section 3, we construct two-dimensional difference operators $L_{\varepsilon,\delta}$ that converge to two-dimensional finite-gap Schr\"odinger operators at one energy level (Theorem 3).

\section{Two-Dimensional Finite-Gap Schr\"odinger Operators at One Energy Level}

Before discussing two-dimensional finite-gap Schr\"odinger operators at one energy level and two-dimensional difference operators, we recall some results from \cite{MM} concerning one-dimensional finite-gap Schr\"odinger operators and related one-dimensional difference operators.

\subsection{One-Dimensional Finite-Gap Schr\"odinger Operators}

Let $L_2 = \partial_x^2 + u(x)$ be a finite-gap Schr\"odinger operator commuting with some operator
$$
L_{2g+1}=\partial_x^{2g+1}+u_{2g}(x)\partial_x^{2g}+\ldots+u_0(x).
$$
The spectral curve of these operators is a hyperelliptic curve $\Gamma$ defined by the equation
$$
w^2=z^{2g+1}+c_{2g}z^{2g}+\dots+c_0.
$$

The common eigenfunction (Baker--Akhiezer function) $\phi$ of the operators
\[
L_2 \phi = z \phi, \quad L_{2g+1} \phi = w \phi, \quad \phi = \phi(x, P), \quad P = (z, w) \in \Gamma,
\]
has an essential singularity at the infinite point $\infty$ of the spectral curve $\Gamma$ and simple poles at certain points $\gamma_1, \ldots, \gamma_g \in \Gamma$ (see \cite{N}--\cite{DMN}).

In \cite{MM}, it was proved that there exists a difference operator of the form
\[
\tilde{L}_2 = \frac{T_\varepsilon^2}{\varepsilon^2} + A(x, \varepsilon) \frac{T_\varepsilon}{\varepsilon} + B(x, \varepsilon),
\]
which commutes with a difference operator $\tilde{L}_{2g+1}$ of order $2g + 1$. The spectral curve of $\tilde{L}_2$ and $\tilde{L}_{2g+1}$ coincides with $\Gamma$ (and is independent of $\varepsilon$). The difference operators are constructed from the points $\gamma_1, \ldots, \gamma_g$ and a family of points $p(x, \varepsilon) \in \Gamma$. Moreover, if $p \to \infty$ as $\varepsilon \to 0$, then the operators $\tilde{L}_2, \tilde{L}_{2g+1}$ converge to the differential operators $L_2, L_{2g+1}$. Note that the maximal commutative ring of differential operators containing $L_2, L_{2g+1}$ is isomorphic to the maximal commutative ring of difference operators containing $\tilde{L}_2, \tilde{L}_{2g+1}$. Thus, the operator $L_2$ can be extended to a difference operator $\tilde{L}_2$ while preserving the fundamental integrable properties. However, this extension is not unique and depends on the choice of the family of points $p(x, \varepsilon) \in \Gamma$.

\subsection{Two-Point Baker--Akhiezer Function}

The two-point Baker--Akhiezer function $\varphi$ (see \cite{DKN}), which lies in the kernel of a two-dimensional finite-gap at one energy level Schr\"odinger operator $H$, is constructed from the following spectral data
$$
S=\{\Gamma,\gamma,p,q\},
$$
where $\gamma=\gamma_1+\ldots+\gamma_g$ is a non-special divisor of degree $g$ on a Riemann surface $\Gamma$ of genus $g$, and $p,q\in\Gamma$ are two marked points.

\noindent{\bf 1}. In neighbourhoods of $p$ and $q$, the function $\varphi(z,\bar{z},P), \ P \in \Gamma$ has the form
$$
\varphi=e^{zk_1}\left(1+\frac{\xi(z,\bar{z})}{k_1}+\ldots\right),
$$
$$
\varphi=e^{\bar{z}k_2}\left(c(z,\bar{z})+\frac{\eta(z,\bar{z})}{k_2}+\ldots\right),
$$
where $k_1^{-1}$, $k_2^{-1}$ are local parameters on $\Gamma$ in neighbourhoods of $p$ and $q$ respectively.

\noindent{\bf 2}. On $\Gamma \backslash \{p, q\}$ the function $\varphi$ is meromorphic with a pole divisor $\gamma$.

The function $\varphi$ is expressed explicitly in terms of the theta function of the Jacobian variety of the surface $\Gamma$.

Let $a_j, b_j, \ j = 1,\ldots,g$ be a basis of cycles on $\Gamma$ with intersection indices
$$
a_i \circ a_j = b_i \circ b_j = 0, \quad a_i \circ b_j = \delta_{ij}.
$$
Denote by $\omega_1,\ldots,\omega_g$ the basis of normalized Abelian differentials $\oint\limits_{a_i} \omega_j = \delta_{ij}.$ The Jacobian variety $J(\Gamma)$ of the surface $\Gamma$ has the form
$$
J(\Gamma) = \mathbb{C}^g/\{\mathbb{Z}^g+\Omega \mathbb{Z}^g\}, \quad \Omega_{ij} = \Omega_{ji} = \oint\limits_{b_i} \omega_j.
$$
The theta function of the Jacobian variety $J(\Gamma)$ is given by the series
$$
\theta(z) = \sum_{n \in \mathbb{Z}^g} \exp\big(\pi i n^t\Omega n+ 2\pi i n^tz \big).
$$
The theta function has the following periodicity properties
$$
\theta(z + n) = \theta(z), \quad n \in \mathbb{Z}^g,
$$
$$
\theta(z + \Omega m) = \exp\big(- \pi i m^t\Omega m - 2\pi i m^tz \big) \theta(z), \quad m \in \mathbb{Z}^g.
$$

The Abel map ${\cal A}: \Gamma \rightarrow J(\Gamma)$ is defined as follows
$$
{\cal A}(P) = \bigg(\int_{P_0}^{P} \omega_1,\ldots,\int_{P_0}^{P} \omega_g\bigg),
$$
where $P_0 \in \Gamma$ is some fixed point. Let $\zeta = -{\cal A}(\gamma)-\mathcal{K} = -{\cal A}(\gamma_1) - \ldots - {\cal A}(\gamma_g) - \mathcal{K},$ where $\mathcal{K}$ is the vector of Riemann constants. Denote by $\Omega_{p},$ $\Omega_{q}$ the meromorphic 1-forms on $\Gamma$ with second-order poles at points $p$ and $q$ respectively, normalized by the conditions $\oint\limits_{a_i} \Omega_{p} = 0, \ \oint\limits_{a_i} \Omega_{q} = 0,$ and let $U^p, \ V^q$ be the vectors of $b$-periods of the differentials $\Omega_{p},$ $\Omega_{q},$
\begin{equation} 
\label{u4}
U^p = \cfrac{1}{2 \pi i} \left(\oint\limits_{b_1} \Omega_p, \ldots,  \oint\limits_{b_g} \Omega_p \right), \qquad 
V^q = \cfrac{1}{2 \pi i} \left( \oint\limits_{b_1} \Omega_q, \ldots, \oint\limits_{b_g} \Omega_q \right).
\end{equation}
Then the function $\varphi$ has the form
$$
\varphi = \exp\bigg(z \Big(\int_{P_0}^{P}\Omega_{p} - \alpha_p\Big) + \bar{z} \Big(\int_{P_0}^{P}\Omega_{q} - \beta_q\Big)\bigg)
\frac{\theta({\cal A}(P)+zU^p+\bar{z}V^q+\zeta)\theta({\cal A}(p)+\zeta)}{\theta({\cal A}(P)+\zeta)\theta({\cal A}(p)+zU^p+\bar{z}V^q+\zeta)},
$$
where $\beta_q = \int\limits_{P_0}^p \Omega_{q},$ and the constant $\alpha_p$ is such that $\int\limits_{P_0}^{P} \Omega_p - \alpha_p = k_1 + O(k_1^{-1}),$ as $P \to p.$

From the existence and uniqueness, it follows that $\varphi$ satisfies the equation
$$
H\varphi= (\partial_z \partial_{\bar{z}} + A(z,\bar{z}) \partial_{\bar{z}} + u(z,\bar{z})) \varphi=0,
$$
where 
$$
A(z,\bar{z}) = -\partial_z\ln c(z,\bar{z}) = -\partial_z\ln \bigg(\frac{\theta ({\cal A}(q) +z U^p +\bar{z}V^q+ \zeta)}{\theta ({\cal A}(p) +z U^p +\bar{z}V^q + \zeta)} \bigg)+const,
$$
$$
u(z,\bar{z}) = -\partial_{\bar{z}} \xi(z,\bar{z}) = \partial_z\partial_{\bar{z}} \ln \theta ({\cal A}(p) +z U^p +\bar{z}V^q + \zeta) + const.
$$

{\bf Remark 1}
Later (Theorem 3), we will need a finite-gap at one energy level Schr\"odinger operator, which is obtained from $H$ by a gauge transformation:
$$
e^{c_1 z} H e^{-c_1 z} =\partial_z \partial_{\bar{z}} + \tilde{A}(z,\bar{z}) \partial_{\bar{z}} + u(z,\bar{z}),
$$
where 
$$
\tilde{A}(z,\bar{z})=A(z,\bar{z})-c_1.
$$
The function $e^{c_1 z}\varphi$ lies in the kernel of the operator $e^{c_1 z} H e^{-c_1 z}.$
We will denote the new Schr\"odinger operator by the same symbol $H$.

\noindent{\bf Example 1}. For $g=1$, the function $\varphi$ and the coefficients of the operator $H$ can be expressed in terms of the Weierstrass elliptic functions $\sigma(w), \zeta(w), \wp(w).$ Let $\Gamma = \mathbb{C} / \{ 2 m \omega + 2 n \omega', n,m \in \mathbb Z \}$ be an elliptic curve. The Weierstrass $\wp$-function is a meromorphic function on $\Gamma$ with a single second-order pole at $0 \in \Gamma,$ defined by the series
$$
\wp (w) = \cfrac{1}{w^2} + \sum\limits_{(n,m) \in \mathbb Z ^2 \backslash \{ 0, 0\} } \left( \cfrac{1}{(w - 2 m \omega - 2 n \omega')^2} - \cfrac{1}{(2 m \omega + 2 n \omega' )^2} \right).
$$
The function $\zeta(w)$ has simple poles at points $w = 2 m \omega + 2 n \omega', n,m \in \mathbb Z$ and satisfies the following identities
$$
\zeta'(w) = - \wp(w), \qquad	\zeta (w + 2 n \omega + 2 m \omega') = \zeta (w) + 2 n \zeta (\omega) + 2 m \zeta (\omega'). 
$$
The function $\sigma(w)$ is an entire function on $\mathbb C,$ with simple zeros at points $w = 2 m \omega + 2 n \omega', n,m \in \mathbb Z$ and satisfies the identities
$$
\cfrac{\sigma'(w)}{\sigma(w)} = \zeta(w), \qquad	\sigma(w + 2 \omega) = - \sigma (w) \exp{ (2 \eta (w + \omega)) }, \qquad \sigma(w + 2 \omega') = - \sigma(w) \exp{ (2 \eta' (w + \omega')) }.
$$
Let $\gamma, p, q \in \Gamma,$ then the finite-gap at one energy level Schr\"odinger operator has the form
$$
H = \partial_z\partial_{\bar{z}} + (\zeta (p - q) + \zeta (q - z - \bar{z} - \gamma) - \zeta (p - z - \bar{z} - \gamma) )  \partial_{\bar{z}} + \notag \\ + \wp (p - q) - \wp (p - z - \bar{z} - \gamma).
$$
The two-point Baker--Akhiezer function has the form
$$
\varphi(z, \bar{z}, w) = e^{z \zeta (w - p) + \bar{z} \zeta (w - q)} \cfrac{\sigma (w - z - \bar{z} - \gamma) \sigma (p - \gamma) }{ \sigma (w - \gamma) \sigma (p - z - \bar{z} - \gamma)} e^{- \bar{z} \zeta (p - q)}.
$$

\subsection{Two-Dimensional Difference Operators}

Let us recall the construction of the difference operator $L$ of the form (\ref{eq2}) from \cite{LM}.

Consider spectral data of the form
$$
\tilde{S} = \{ \Gamma, \gamma, p, q, p_n, q_m \}, \quad n,m \in \mathbb{Z},
$$
where $\Gamma$ is a compact Riemann surface of genus $g$, $p, q$ are two marked points on $\Gamma$, $\gamma = \gamma_1 + \ldots + \gamma_g$ is a non-special divisor of degree $g$ on $\Gamma$, and $p_n, q_m$ are two families of points on $\Gamma$ in general position.

Let us introduce the following divisors:
$$
P(n) =
\begin{cases}
p_1 + \ldots + p_n, & n > 0 \\
-p_0 - \ldots - p_{n+1}, & n < 0, \\
0, & n = 0
\end{cases}
\quad
Q(m) =
\begin{cases}
q_1 + \ldots + q_m, & m > 0 \\
-q_0 - \ldots - q_{m+1}, & m < 0. \\
0, & m = 0
\end{cases}
$$

The following theorem holds.

\begin{theorem} (\cite{LM})
There exists a unique meromorphic function $\psi(n,m,P)$ on $\Gamma$, $n,m\in\mathbb{Z}$, $P\in\Gamma$, satisfying the following conditions:

\noindent{\bf 1}. The zero and pole divisor of $\psi$ has the form
$$
\big( \psi(n,m,P) \big) = P(n) + Q(m) + \gamma_{n,m} - n p - m q - \gamma,
$$
where $\gamma_{n,m} = \gamma_1(n,m) + \ldots + \gamma_g(n,m)$ is some divisor on $\Gamma$, with $\gamma_{0,0} = \gamma$.

\noindent{\bf 2}. In a neighbourhood of $p$, the function $\psi(n,m,P)$ has the form
$$
\psi(n,m,P) = k_1^n + O(k_1^{n-1}),
$$
where $k_1^{-1}$ is a local parameter in the neighbourhood of $p$.

\noindent{\bf 3}. $\psi(0,0,P) = 1$.

Moreover, the function $\psi(n,m,P)$ satisfies the equation
$$
L\psi(n,m,P) = \psi(n+1,m+1,P) + a_{n,m}\psi(n+1,m,P) + b_{n,m}\psi(n,m+1,P) + v_{n,m}\psi(n,m,P) = 0,
$$
where $a_{n,m}, b_{n,m}, v_{n,m}$ are some coefficients.
\end{theorem}

Thus, the kernel of the operator $L$ contains a family of functions parameterized by points of the Riemann surface $\Gamma$.

In what follows, we will need a certain generalization of the function $\psi(n,m,P)$. Let $\varepsilon, \delta \in \mathbb{C}$. Consider the following spectral data:
$$
\tilde{S}_{\varepsilon,\delta} = \{\Gamma, \gamma, p, q, p(\varepsilon n), q(\delta m) \}, \quad n,m \in \mathbb{Z},
$$
where $p(\varepsilon n), q(\delta m) \in \Gamma$ are two families of points in general position (it is convenient for us to use the notation $\varepsilon n, \delta m$ for indices and function arguments rather than $n,m$, since later $\varepsilon n, \delta m$ will be extended to complex $z_1,z_2 \in \mathbb{C}$).

Let us introduce the divisors:
$$
P(\varepsilon n) =
\begin{cases}
p_{\varepsilon} + \ldots + p_{\varepsilon n}, & n > 0 \\
-p_{0} - \ldots - p_{\varepsilon (n+1)}, & n < 0, \\
0, & n = 0
\end{cases}
\quad
Q(\delta m) =
\begin{cases}
q_{\delta} + \ldots + q_{\delta m}, & m > 0 \\
-q_{0} - \ldots - q_{\delta (m+1)}, & m < 0. \\
0, & m = 0
\end{cases}
$$

There exists a unique meromorphic function $\psi_{\varepsilon,\delta}(\varepsilon n,\delta m,P)$ on $\Gamma$, $P \in \Gamma$, which has the following properties:

\noindent{\bf 1}. The zero and pole divisor of $\psi_{\varepsilon,\delta}$ has the form
\begin{equation}
    \label{eq4}
    \big( \psi_{\varepsilon,\delta}(\varepsilon n,\delta m, P) \big) = P(\varepsilon n) + Q(\delta m) + \gamma(\varepsilon n,\delta m) - n p - m q - \gamma,
\end{equation}
where $\gamma(\varepsilon n,\delta m) = \gamma_1(\varepsilon n,\delta m) + \ldots + \gamma_g(\varepsilon n,\delta m)$, with $\gamma(0,0) = \gamma$.

\noindent{\bf 2}. In a neighbourhood of the point $p$, the function $\psi_{\varepsilon,\delta}$ has the form
\begin{equation}
    \label{eq5}
    \psi_{\varepsilon,\delta}(\varepsilon n,\delta m, P) = (\varepsilon k_1)^n + O(k_1^{n-1}).
\end{equation}

\noindent{\bf 3}. $\psi_{\varepsilon,\delta}(0,0,P) = 1$.

The only essential difference between the function $\psi_{\varepsilon,\delta}(\varepsilon n,\delta m,P)$ and the function $\psi(n,m,P)$ in Theorem 1 lies in the asymptotic expansion (\ref{eq5}).

The function $\psi_{\varepsilon,\delta}$ satisfies the equation
$$ 
\tilde{L}_{\varepsilon,\delta} \psi_{\varepsilon,\delta}(\varepsilon n,\delta m,P) = \Big(\frac{T_{\varepsilon}}{\varepsilon}\frac{T_{\delta}}{\delta} + a(\varepsilon n,\delta m) \frac{T_{\varepsilon}}{\varepsilon} + b(\varepsilon n,\delta m) \frac{T_{\delta}}{\delta} + v(\varepsilon n,\delta m)\Big) \psi_{\varepsilon,\delta}(\varepsilon n,\delta m,P) = 0,
$$
where $a(\varepsilon n,\delta m), b(\varepsilon n,\delta m), v(\varepsilon n,\delta m)$ are some functions. The proof of the existence of $\psi_{\varepsilon,\delta}$ and of the relation $\tilde{L}_{\varepsilon,\delta} \psi_{\varepsilon,\delta} = 0$ is exactly the same as for Theorem 1 (see \cite{LM}).

Note that in a neighbourhood of $q$, we have an expansion of the form
\begin{equation} 
    \label{eq6}
    \psi(\varepsilon n,\delta m, P) = \lambda(\varepsilon n,\delta m)(\delta k_2)^m + O(k_2^{m-1}), 
\end{equation}
where $\lambda(0,0) = 1$, and $k^{-1}_2$ is a local parameter in the neighbourhood of $q$.

Let us find an explicit formula for $\psi_{\varepsilon,\delta}(\varepsilon n,\delta m,P)$ using the theta function of the Jacobian variety of the surface $\Gamma$.

Let $\Omega_{\varepsilon n}$ be a meromorphic 1-form on $\Gamma$ with two simple poles at points $p(\varepsilon n)$, $p$ and residues $\operatorname{Res}_{p(\varepsilon n)} \Omega_{\varepsilon n} = \frac{1}{\varepsilon}$, $\operatorname{Res}_{p} \Omega_{\varepsilon n} = -\frac{1}{\varepsilon}$, normalized by the condition
\begin{equation} 
    \label{eq7}
    \oint\limits_{a_k} \Omega_{\varepsilon n} = 0, \quad k = 1, \ldots, g.
\end{equation}
Similarly, let $\Delta_{\delta m}$ be a meromorphic 1-form with two simple poles at points $q(\delta m)$, $q$ and residues $\operatorname{Res}_{q(\delta m)} \Delta_{\delta m} = \frac{1}{\delta}$, $\operatorname{Res}_{q} \Delta_{\delta m} = -\frac{1}{\delta}$, normalized by the condition
\begin{equation} 
    \label{eq8}
    \oint\limits_{a_k} \Delta_{\delta m} = 0, \quad k = 1, \ldots, g.
\end{equation}
In what follows, we will need the representation of the form $\Delta_{\delta m}$ in a neighbourhood of $p$. Let $\Delta_{\delta m} = g_{\delta m}(k_1^{-1})dk_1^{-1}$, where $g_{\delta m}$ is some function.

Let us introduce the vectors
$$
U_{\varepsilon n} = \cfrac{\varepsilon}{2 \pi i} \left( \oint\limits_{b_1} \Omega_{\varepsilon n}, \ldots,  \oint\limits_{b_g} \Omega_{\varepsilon n} \right), \qquad 
V_{\delta m} = \cfrac{\delta}{2 \pi i} \left( \oint\limits_{b_1} \Delta_{\delta m}, \ldots,  \oint\limits_{b_g} \Delta_{\delta m}\right).
$$	
Let
$$
\Omega (\varepsilon n) = 
\begin{cases}
\Omega_{\varepsilon} + \ldots + \Omega_{\varepsilon n}, & n > 0 \\
-\Omega_{0} - \ldots -\Omega_{\varepsilon (n+1)}, & n < 0, \\
0, & n = 0
\end{cases} \quad
\Delta(\delta m) = 
\begin{cases}
\Delta_{\delta} + \ldots + \Delta_{\delta m}, & m > 0 \\
- \Delta_{0} - \ldots - \Delta_{\delta (m+1)}, & m < 0. \\
0, & m = 0
\end{cases}
$$
We denote their $b$-period vectors by
$$
U(\varepsilon n) = \cfrac{\varepsilon}{2 \pi i} \left( \oint\limits_{b_1} \Omega(\varepsilon n), \ldots,  \oint\limits_{b_g} \Omega(\varepsilon n) \right) = 
\begin{cases}
U_{\varepsilon} + \ldots + U_{\varepsilon n}, & n > 0 \\
-U_{0} - \ldots -U_{\varepsilon (n+1)}, & n < 0, \\
0, & n = 0
\end{cases} \quad
$$
$$
V(\delta m) = \cfrac{\delta}{2 \pi i} \left( \oint\limits_{b_1} \Delta(\delta m), \ldots,  \oint\limits_{b_g} \Delta(\delta m) \right) = 
\begin{cases}
V_{\delta} + \ldots + V_{\delta m}, & m > 0 \\
- V_{0} - \ldots - V_{\delta (m+1)}, & m < 0. \\
0, & m = 0
\end{cases}
$$

Note that the vectors $U(\varepsilon n)$ and $V(\delta m)$ are solutions of the difference equations
\begin{equation} 
\label{eq9}
\cfrac{\varepsilon}{2 \pi i} \oint\limits_{b} \Omega_{\varepsilon (n+1)} = U(\varepsilon (n+1)) - U(\varepsilon n), 
\qquad
\cfrac{\delta}{2 \pi i} \oint\limits_{b} \Delta_{\delta (m+1)} = V(\delta (m+1)) - V(\delta m).
\end{equation} 
Let, as in the case of the Schr\"odinger operator, $\zeta=-{\cal A}(\gamma) - \mathcal{K}.$ Define the constant $\alpha_{\varepsilon n}$ from the expansion
\begin{equation} 
\label{u11}
\varepsilon \int\limits_{P_0}^{P} \Omega(\varepsilon n) - \alpha_{\varepsilon n} = n \ln{k_1} + n \ln{\varepsilon} + O(k_1^{-1}), \quad \text{as} \quad P \to p.
\end{equation}
Note that from (\ref{u11}) it follows that
\begin{equation} 
\label{u12}
\bigg(\varepsilon \int\limits_{P_0}^{P} \Omega_{\varepsilon(n+1)} - \varepsilon \ln{k_1}-\varepsilon \ln{\varepsilon} \bigg)\bigg|_{P=p} = \alpha_{\varepsilon(n+1)} - \alpha_{\varepsilon n}.
\end{equation}
Let $\beta_{\delta m} = \delta\int\limits_{P_0}^p \Delta(\delta m).$

The following theorem holds:
\begin{theorem}	
    The function
    \begin{equation} 
        \label{eq10}
        \psi_{\varepsilon,\delta}(\varepsilon n, \delta m,P) = \exp{ \left( \varepsilon \int\limits_{P_0}^{P} \Omega(\varepsilon n) - \alpha_{\varepsilon n} + \delta \int\limits_{P_0}^{P} \Delta(\delta m) - \beta_{\delta m} \right)}
        \cfrac{\theta ({\cal A}(P) + U(\varepsilon n) + V(\delta m) + \zeta) \theta ({\cal A}(p) + \zeta)}{\theta ({\cal A}(p) + U(\varepsilon n) + V(\delta m) + \zeta)\theta ({\cal A}(P) + \zeta)},
    \end{equation} 
    is a solution to the equation $\tilde{L}_{\varepsilon,\delta} \psi_{\varepsilon,\delta} = 0,$ where 
    \begin{equation} 
    \label{eq11}
    \tilde{L}_{\varepsilon,\delta}= \frac{T_{\varepsilon}}{\varepsilon}\frac{T_{\delta}}{\delta} + a(\varepsilon n,\delta m) \frac{T_{\varepsilon}}{\varepsilon} + b(\varepsilon n,\delta m) \frac{T_{\delta}}{\delta} + v(\varepsilon n,\delta m).
    \end{equation} 
    The coefficients of the operator $\tilde{L}_{\varepsilon,\delta}$ have the form
    $$
    a(\varepsilon n,\delta m) = -\frac{1}{\delta}, 
    $$
    $$
    b(\varepsilon n,\delta m) = 
    - \frac{1}{\varepsilon}\exp{\left( \varepsilon \int\limits_{P_0}^q \Omega_{\varepsilon (n+1)} - \alpha_{\varepsilon (n+1)} + \alpha_{\varepsilon n} \right)} \times
    $$
    $$
    \cfrac{\theta ({\cal A}(q) + U(\varepsilon(n+1)) + V(\delta(m+1)) + \zeta) \theta ({\cal A}(p) + U(\varepsilon n) + V(\delta (m+1)) + \zeta)}{\theta ({\cal A}(p) + U(\varepsilon(n+1)) + V(\delta(m+1)) + \zeta)\theta ({\cal A}(q) + U(\varepsilon n) + V(\delta(m+1)) + \zeta) },
    $$
    $$
    \left. v(\varepsilon n,\delta m) = -\frac{1}{\delta} \bigg(b(\varepsilon n,\delta m) + \delta g_{\delta (m+1)}(0) + \cfrac{\partial}{\partial k_1^{-1}} \ln \bigg(\cfrac{\theta({\cal A}(P) + U(\varepsilon(n+1)) + V(\delta (m+1)) + \zeta)}{\theta ({\cal A}(P) + U(\varepsilon(n+1)) + V(\delta m) + \zeta)} \bigg)\right\vert_{P = p} \bigg).
    $$
\end{theorem}
In formula (\ref{eq10}), it is assumed that the path from $P_0$ to $P$ in the integrals $\int\limits_{P_0}^{P} \Omega(\varepsilon n), \ \int\limits_{P_0}^{P} \Delta(\delta m)$ and in the Abel map ${\cal A}(P)$ is the same.

{\bf Proof.} 
From the properties of the theta function, conditions (\ref{eq7}), (\ref{eq8}), and the definition of the vectors $U(\varepsilon n), V(\delta m)$, it follows that the value of the function $\psi_{\varepsilon,\delta}(\varepsilon n, \delta m,P)$ does not depend on the choice of the path from $P_0$ to $P$. Consequently, $\psi_{\varepsilon,\delta}$ is well-defined on $\Gamma$. From Riemann's theorem on zeros of the theta function and the definition of $\Omega(\varepsilon n)$ and $\Delta(\delta m)$, it follows that the divisor of zeros and poles of $\psi_{\varepsilon,\delta}$ has the form (\ref{eq4}), and the expansion (\ref{eq5}) holds. Moreover, for $n=m=0$ we have $\psi(0,0,P)=1$.

Let us find the coefficients of the operator $\tilde{L}_{\varepsilon,\delta}.$
For this purpose, we will need the functions
\begin{equation} 
    \label{eq12}
\chi_1(\varepsilon n,\delta m, P)=\frac{\psi_{\varepsilon,\delta}(\varepsilon (n+1),\delta m,P)}{\varepsilon \psi_{\varepsilon,\delta}(\varepsilon n,\delta m,P)}, \qquad 
\chi_2(\varepsilon n,\delta m, P)=\frac{\psi_{\varepsilon,\delta}(\varepsilon n,\delta (m+1),P)}{\delta \psi_{\varepsilon,\delta}(\varepsilon n,\delta m,P)}.
\end{equation}
From (\ref{eq5}), (\ref{eq6}), and (\ref{eq12}), the following expansions hold in a neighbourhood of $p$:
$$
\chi_1(\varepsilon n,\delta m, P) = k_1 + d_0(\varepsilon n,\delta m) + O(k^{-1}_1), \qquad
\chi_2(\varepsilon n,\delta m, P) = \frac{1}{\delta} + \frac{c_1(\varepsilon n,\delta m)}{k_1} + O(k_1^{-2})
$$
and in a neighbourhood of $q$:
$$
\chi_1(\varepsilon n,\delta m, P) = s_0 (\varepsilon n,\delta m) + \frac{s_1 (\varepsilon n,\delta m)}{k_2} + O(k_2^{-2}), \qquad
\chi_2(\varepsilon n,\delta m, P) = u(\varepsilon n,\delta m)k_2 + u_0(\varepsilon n,\delta m) + O(k_2^{-1}),
$$
where $d_0(\varepsilon n,\delta m), c_1(\varepsilon n,\delta m), s_0(\varepsilon n,\delta m), s_1(\varepsilon n,\delta m), u(\varepsilon n,\delta m), u_0 (\varepsilon n,\delta m)$ are some functions. The coefficients of the operator $\tilde{L}_{\varepsilon,\delta}$ are expressed via the coefficients $c_1(\varepsilon n,\delta m)$ and $s_0 (\varepsilon n,\delta m)$ by the formulas (the calculations are exactly the same as for the coefficients of the operator $L$ (see \cite{LM}))
$$
a(\varepsilon n,\delta m) =  - \chi_2 (\varepsilon (n+1), \delta m, p) = -\frac{1}{\delta}, 
$$
$$
b(\varepsilon n,\delta m) = - \chi_1 (\varepsilon n,\delta (m + 1), q) = - s_0(\varepsilon n,\delta (m+1)),
$$
$$
v(\varepsilon n,\delta m) = a(\varepsilon (n-1),\delta m) b(\varepsilon n,\delta m) - \cfrac{\partial}{\partial k_1^{-1}} \chi_2 (\varepsilon(n+1), \delta m, P) \Big\vert_{P = p} = \frac{1}{\delta}s_0(\varepsilon n,\delta (m+1))  - c_1(\varepsilon (n+1),\delta m).
$$

Next, let us find the coefficients of the operator $\tilde{L}_{\varepsilon,\delta}$ in terms of the theta function.

From formulas (\ref{eq10}) and (\ref{eq12}), we find $\chi_1 (\varepsilon n,\delta m,P)$ and $\chi_2 (\varepsilon n,\delta m,P).$ 
$$
\chi_1 = \frac{1}{\varepsilon}\exp{\left( \varepsilon \int\limits_{P_0}^P \Omega_{\varepsilon (n+1)} - \alpha_{\varepsilon (n+1)} + \alpha_{\varepsilon n} \right)} \cfrac{\theta ({\cal A}(P) + U(\varepsilon (n+1)) + V(\delta m) + \zeta) \theta ({\cal A}(p) + U(\varepsilon n) + V(\delta m) + \zeta)}{\theta ({\cal A}(p) + U(\varepsilon (n+1)) + V(\delta m) + \zeta)\theta ({\cal A}(P) + U(\varepsilon n) + V(\delta m) + \zeta)},  
$$
$$
\chi_2= \frac{1}{\delta}\exp{\left( \delta \int\limits_{P_0}^P \Delta_{\delta(m+1)} - \beta_{\delta(m+1)} + \beta_{\delta m} \right)} \cfrac{\theta ({\cal A}(P) + U(\varepsilon n) + V(\delta(m+1)) + \zeta) \theta ({\cal A}(p) + U(\varepsilon n) + V(\delta m) + \zeta)}{\theta ({\cal A}(p) + U(\varepsilon n) + V(\delta(m+1)) + \zeta)\theta ({\cal A}(P) + U(\varepsilon n) + V(\delta m) + \zeta)}.
$$

From the definition of $\beta_{\delta m}$ and $\Delta(\delta m)$, it follows that
$$
\exp{\left( \delta \int\limits_{P_0}^p \Delta_{\delta(m+1)} - \beta_{\delta(m+1)} + \beta_{\delta m} \right)} = \exp{\left( \delta \int\limits_{P_0}^p \Delta_{\delta(m+1)} -  \delta\int\limits_{P_0}^p \Delta(\delta (m+1)) +  \delta\int\limits_{P_0}^p \Delta(\delta m) \right)} = 1.
$$
Consequently,
$$
a(\varepsilon n,\delta m) =  - \chi_2 (\varepsilon (n+1), \delta m, p) = -\frac{1}{\delta}. 
$$
The remaining coefficients of the operator $L_{\varepsilon,\delta}$ have the form
$$
b(\varepsilon n,\delta m) = - \chi_1 (\varepsilon n,\delta (m + 1), q) =
- \frac{1}{\varepsilon}\exp{\left( \varepsilon \int\limits_{P_0}^q \Omega_{\varepsilon (n+1)} - \alpha_{\varepsilon (n+1)} + \alpha_{\varepsilon n} \right)} \times
$$
$$
\cfrac{\theta ({\cal A}(q) + U(\varepsilon(n+1)) + V(\delta(m+1)) + \zeta) \theta ({\cal A}(p) + U(\varepsilon n) + V(\delta (m+1)) + \zeta)}{\theta ({\cal A}(p) + U(\varepsilon(n+1)) + V(\delta(m+1)) + \zeta)\theta ({\cal A}(q) + U(\varepsilon n) + V(\delta(m+1)) + \zeta) },
$$
$$
v(\varepsilon n,\delta m)  = -\frac{1}{\delta} b(\varepsilon n,\delta m) - \cfrac{\partial}{\partial k_1^{-1}} \chi_2 (\varepsilon(n+1), \delta m, P) \Big\vert_{P = p} =
$$
$$
=-\frac{1}{\delta} b(\varepsilon n,\delta m) - \frac{1}{\delta}\cfrac{\partial}{\partial k_1^{-1}} \exp{\left( \delta \int\limits_{P_0}^P \Delta_{\delta(m+1)} - \beta_{\delta(m+1)} + \beta_{\delta m} \right)} \bigg\vert_{P = p} - 
$$
$$
-\frac{1}{\delta}\cfrac{\theta ({\cal A}(p) + U(\varepsilon (n+1)) + V(\delta m) + \zeta)}{\theta ({\cal A}(p) + U(\varepsilon (n+1)) + V(\delta(m+1)) + \zeta)} \times 
\cfrac{\partial}{\partial k_1^{-1}} \cfrac{\theta ({\cal A}(P) + U(\varepsilon (n+1)) + V(\delta(m+1)) + \zeta)}{\theta ({\cal A}(P) + U(\varepsilon (n+1)) + V(\delta m) + \zeta)} \bigg\vert_{P = p} =
$$
$$
= -\frac{1}{\delta} \bigg(b(\varepsilon n,\delta m) + \delta g_{\delta(m+1)}(0) + \cfrac{\partial}{\partial k_1^{-1}} \ln \bigg(\cfrac{\theta({\cal A}(P) + U(\varepsilon(n+1)) + V(\delta (m+1)) + \zeta)}{\theta ({\cal A}(P) + U(\varepsilon(n+1)) + V(\delta m) + \zeta)} \bigg)\bigg\vert_{P = p} \bigg).
$$
Theorem 2 is proved.

\section{Two-Dimensional Difference Operators}

In this section, we will extend the operator $\tilde{L}_{\varepsilon,\delta}$ (see (\ref{eq11})) to a difference operator $L_{\varepsilon,\delta}$ of the form (\ref{eq3}), which will have a limit as $\varepsilon,\delta \to 0.$

Let us modify the spectral data $\tilde{S}_{\varepsilon,\delta}$ to the following spectral data
$$
S_{\varepsilon,\delta} = \{\Gamma, \gamma, p, q, p(z_1,\varepsilon), q(z_2,\delta) \}, \quad z_1,z_2 \in \mathbb{C}, 
$$
where $p(z_1,\varepsilon), q(z_2,\delta) \in \Gamma$ are smooth families of points such that $p(z_1,0) = p$, $q(z_2,0) = q$. We will assume that their coordinates have the form
\begin{equation} 
    \label{u16}
k^{-1}_1(p(z_1,\varepsilon)) = -\varepsilon + O(\varepsilon^2), \qquad k^{-1}_2(q(z_2,\delta)) = -\delta + O(\delta^2).
\end{equation}

Let $\Omega_{z_1}$ denote the meromorphic 1-form on $\Gamma$ with poles at points $p(z_1,\varepsilon)$, $p$ and residues $\operatorname{Res}_{p(z_1,\varepsilon)} \Omega_{z_1} = \frac{1}{\varepsilon}$, $\operatorname{Res}_{p} \Omega_{z_1} = -\frac{1}{\varepsilon}$. 
Similarly, let $\Delta_{z_2}$ denote the meromorphic 1-form on $\Gamma$ with poles at points $q(z_2,\delta)$, $q$ and residues $\operatorname{Res}_{q(z_2,\delta)} \Delta_{z_2} = \frac{1}{\delta}$, $\operatorname{Res}_{q} \Delta_{z_2} = -\frac{1}{\delta}$. 
Let $\Omega_{p}$ and $\Omega_{q}$, as in Section 2.2, denote meromorphic 1-forms with second-order poles at points $p$, $q$ respectively, having expansions of the form
$$
\Omega_{p} = \Big(-\frac{1}{k^{-2}_1}+O(1)\Big)dk^{-1}_1, \qquad 
\Omega_{q} = \Big(-\frac{1}{k^{-2}_2}+O(1)\Big)dk^{-1}_2.
$$
Note that $\lim\limits_{\varepsilon \to 0} \Omega_{z_1} = \Omega_p$ (similarly $\lim\limits_{\delta \to 0} \Delta_{z_2} = \Omega_q$). Indeed, $\oint\limits_{a_k} \Omega_{z_1} = \oint\limits_{a_k} \Omega_{p} = 0, \quad k = 1, \ldots, g.$ In a neighbourhood of $p$, by (\ref{u16}), we have the expansion
$$
\Omega_{z_1} = \bigg(\frac{1}{\varepsilon(k^{-1}_1-k^{-1}_1(p(z_1,\varepsilon)))}-\frac{1}{\varepsilon k^{-1}_1} + O(1)\bigg)dk^{-1}_1 = \bigg(\frac{-1+O(\varepsilon)}{(k^{-1}_1-k^{-1}_1(p(z_1,\varepsilon)))k^{-1}_1} + O(1)\bigg)dk^{-1}_1.
$$
Consequently, in a neighbourhood of $p$,
$$
\lim\limits_{\varepsilon \to 0} \Omega_{z_1}= \bigg(-\frac{1}{k^{-2}_1} + O(1)\bigg)dk^{-1}_1.
$$
Thus, the form $(\lim\limits_{\varepsilon \to 0} \Omega_{z_1} - \Omega_p)$ is holomorphic and has zero $a$-periods, which means this form is identically zero form.

In a neighbourhood of the point $p$, the form $\Delta_{z_2}$ has the form
$\Delta_{z_2} = g_{z_2}(k_1^{-1})dk_1^{-1}$ (the function $g_{z_2}$ will be needed later).

Consider the analogues of equations (\ref{eq9}):
\begin{equation} 
    \label{u17}
\cfrac{\varepsilon}{2 \pi i} \oint\limits_{b} \Omega_{z_1 + \varepsilon} = U(z_1 + \varepsilon,\varepsilon) - U(z_1,\varepsilon), \qquad  U(0,\varepsilon) = 0,
\end{equation} 
\begin{equation} 
    \label{u18}
\cfrac{\delta}{2 \pi i} \oint\limits_{b} \Delta_{z_2 + \delta} = V(z_2 + \delta,\delta) - V(z_2,\delta), \qquad V(0, \delta) = 0.
\end{equation}
Note that 
$$
\cfrac{\varepsilon}{2 \pi i} \oint\limits_{b} \Omega_{z_1 + \varepsilon} = \cfrac{\varepsilon}{2 \pi i} \bigg(\oint\limits_{b} \Omega_{p} +O(\varepsilon)\bigg) = \varepsilon U^p + O(\varepsilon^2), 
$$
$$
\cfrac{\delta}{2 \pi i} \oint\limits_{b} \Delta_{z_2 + \delta} = \cfrac{\delta}{2 \pi i} \bigg(\oint\limits_{b} \Omega_{q} +O(\delta)\bigg) = \delta V^q + O(\delta^2),
$$
where the vectors $U^p,V^q$ are defined in (\ref{u4}). The solutions of equations (\ref{u17}), (\ref{u18}) have expansions
\begin{equation} 
    \label{u19}
U(z_1,\varepsilon) = z_1 U^p - \tilde{U}(z_1) \varepsilon + O(\varepsilon^2), \qquad V(z_2,\delta) = z_2 V^q - \tilde{V}(z_2) \delta + O(\delta^2).
\end{equation}

Let us define the function $\alpha_{z_1}$, dependent on $z_1$ and $\varepsilon$ (an analogue of $\alpha_{\varepsilon n}$ in (\ref{u11}) and (\ref{u12})), as the solution to the equation
$$
\bigg(\varepsilon \int\limits_{P_0}^{P} \Omega_{z_1+\varepsilon} - \varepsilon \ln{k_1} -  \varepsilon \ln{\varepsilon}\bigg)\bigg|_{P=p} =\alpha_{z_1+\varepsilon} - \alpha_{z_1}.
$$ 
Then for $z_1=\varepsilon n$, the solution of this equation will coincide with the solution $\alpha_{\varepsilon n}$ of equation (\ref{u12}). Note that
$$
\exp{\left( \varepsilon \int\limits_{P_0}^q \Omega_{z_1+\varepsilon} - \alpha_{z_1+\varepsilon} + \alpha_{z_1} \right)} = 1 + s \varepsilon + O(\varepsilon^2),
$$
where $s$ is some constant. Now, using the formulas for the coefficients of the operator (\ref{eq11}), we define the difference operator $L_{\varepsilon,\delta}$
$$
L_{\varepsilon,\delta} = \frac{T_{\varepsilon}}{\varepsilon} \frac{T_{\delta}}{\delta} + a(z_1,z_2) \frac{T_{\varepsilon}}{\varepsilon} + b(z_1,z_2) \frac{T_{\delta}}{\delta} + v(z_1,z_2),
$$
where
$$
a(z_1,z_2) = -\frac{1}{\delta}, 
$$
$$
b(z_1,z_2) = 
- \frac{1}{\varepsilon}\exp{\left( \varepsilon \int\limits_{P_0}^q \Omega_{z_1+\varepsilon} - \alpha_{z_1+\varepsilon} + \alpha_{z_1} \right)} \times
$$
$$
\cfrac{\theta ({\cal A}(q) + U(z_1+\varepsilon) + V(z_2+\delta) + \zeta) \theta ({\cal A}(p) + U(z_1) + V(z_2+\delta) + \zeta)}{\theta ({\cal A}(q) + U(z_1) + V(z_2+\delta) + \zeta)\theta ({\cal A}(p) + U(z_1+\varepsilon) + V(z_2+\delta) + \zeta)},
$$
$$
\left. v(z_1,z_2)  = -\frac{1}{\delta} \bigg(b(z_1,z_2) + \delta g_{z_2+\delta}(0) + \cfrac{\partial}{\partial k_1^{-1}} \ln \bigg(\cfrac{\theta({\cal A}(P) + U(z_1+\varepsilon) + V(z_2+\delta) + \zeta)}{\theta ({\cal A}(P) + U(z_1+\varepsilon) + V(z_2) + \zeta)} \bigg)\right\vert_{P = p} \bigg).
$$

Note that under the substitution $z_1=\varepsilon n$ and $z_2=\delta m$, the operator $L_{\varepsilon,\delta}$ coincides with $\tilde{L}_{\varepsilon,\delta}$. Let us formulate the main theorem.

\begin{theorem}	
${\ }$

\noindent{\bf 1}. For $z_1=\varepsilon n, \ z_2=\delta m, \ n,m \in \mathbb{Z}$, the kernel of the operator $L_{\varepsilon,\delta}$ contains the discrete Baker-Akhiezer function $\psi_{\varepsilon,\delta}(\varepsilon n,\delta m,P).$

\noindent{\bf 2}.
Let $z_1 = z, \ z_2 = \bar{z}.$ Then as $\varepsilon,\delta \to 0$, the operator $L_{\varepsilon, \delta}$ converges to the finite-gap at one energy level Schr\"odinger operator $H.$
\end{theorem}

{\bf Proof.} 
By construction, the operator $L_{\varepsilon,\delta}$ coincides with $\tilde{L}_{\varepsilon,\delta}$ when $z_1=\varepsilon n, \ z_2=\delta m.$ According to Theorem 2, the kernel of $\tilde{L}_{\varepsilon,\delta}$ contains the discrete Baker-Akhiezer function $\psi_{\varepsilon,\delta}.$ 

Let us prove the second part of Theorem 3. In what follows, we will use the notation $\theta_i(z)=\partial_{z_i}\theta(z).$ The components of the vectors $U^p,V^q,\tilde{U}(z_1),\tilde{V}(z_2)$ in (\ref{u19}) will be denoted as follows:
$$
U^p=(U_1^p,\ldots,U_g^p), \qquad V^q=(V_1^q,\ldots,V_g^q),
$$
$$
\tilde{U}(z_1)=(\tilde{U}_1(z_1),\ldots,\tilde{U}_g(z_1)), \qquad \tilde{V}(z_2)=(\tilde{V}_1(z_2),\ldots,\tilde{V}_g(z_2)).
$$ 
Recall that the Abel map in the neighbourhood of $p$ has the following expansion (see, for example,~\cite{Dub1}):
$$
{\cal A}(P) ={\cal A}(p) - U^p k_1^{-1}+O(k_1^{-2}).
$$

To prove the second part of Theorem 3, we will need expansions of the functions $b(z_1,z_2)$ and $v(z_1,z_2)$ in $\varepsilon$ and $\delta.$ To derive these expansions, we carry out the following calculations. Using (\ref{u19}), we get:
$$
\theta ({\cal A}(q) + U(z_1+\varepsilon) + V(z_2+\delta)  + \zeta) =
\theta({\cal A}(q) + z_1 U^p+ z_2V^q + \zeta) + \varepsilon \sum\theta_i({\cal A}(q) + z_1 U^p+ z_2V^p+ \zeta)(U^p_i- \tilde{U}_i(z_1)) +
$$
$$
+\delta \sum\theta_i({\cal A}(q) + z_1 U^p+ z_2V^p+ \zeta)(V^q_i- \tilde{V}_i(z_2)) + O(\varepsilon^2) + O(\delta^2) + O(\delta\varepsilon).
$$

Here and below, summation is taken over the index $i=1,...,g.$ Similarly, we obtain
$$
\cfrac{\theta ({\cal A}(q) + U(z_1+\varepsilon) + V(z_2+\delta) + \zeta)}{\theta ({\cal A}(q) + U(z_1) + V(z_2+\delta) + \zeta)} = 1 + \varepsilon \frac{\sum\theta_i({\cal A}(q) + z_1 U^p+ z_2V^q+ \zeta)U^p_i}{\theta({\cal A}(q) + z_1 U^p+ z_2V^q+ \zeta)} + O(\varepsilon^2) + O(\delta\varepsilon) =
$$
$$
= 1 + \varepsilon \partial_{z_1}\ln\theta({\cal A}(q) + z_1 U^p+ z_2V^q+ \zeta) + O(\varepsilon^2) + O(\delta\varepsilon),
$$
$$
\cfrac{\theta ({\cal A}(p) + U(z_1) + V(z_2+\delta) + \zeta)}{\theta ({\cal A}(p) + U(z_1+\varepsilon) + V(z_2+\delta) + \zeta)} = 1 - \varepsilon \partial_{z_1}\ln\theta({\cal A}(p) + z_1 U^p+ z_2V^q+ \zeta)+ O(\varepsilon^2) + O(\delta\varepsilon).
$$

Next, we have
$$
b(z_1,z_2)
= \big(- \cfrac{1}{\varepsilon} +s + O(\varepsilon)\big) \bigg(1 + \varepsilon \partial_{z_1}\ln\theta({\cal A}(q) + z_1 U^p+ z_2V^q+ \zeta)  - \varepsilon \partial_{z_1}\ln\theta({\cal A}(p) + z_1 U^p+ z_2V^q+ \zeta)  + 
$$
$$
+O(\varepsilon^2) + O(\delta\varepsilon) \bigg)=
- \cfrac{1}{\varepsilon} - \partial_{z_1}\ln \bigg(\cfrac{\theta({\cal A}(q) + z_1 U^p + z_2 V^q + \zeta)}{\theta ({\cal A}(p) + z_1 U^p + z_2 V^q + \zeta)}\bigg) +s+ O(\varepsilon)+ O(\delta).
$$

Let us find the expansion of the function $v(z_1,z_2)$ in $\varepsilon,\delta$.
$\varepsilon,\delta$
$$
\cfrac{\partial}{\partial k_1^{-1}} \ln \bigg(\cfrac{\theta({\cal A}(P) + U(z_1+\varepsilon) + V(z_2+\delta) + \zeta)}{\theta ({\cal A}(P) + U(z_1+\varepsilon) + V(z_2) + \zeta)} \bigg)\bigg\vert_{P = p} = 
$$
$$
=\cfrac{\partial}{\partial k_1^{-1}} \ln \bigg(1 + \delta\bigg(\cfrac{\sum\theta_i({\cal A}(P)+ z_1U^p + z_2V^q + \zeta)(V_i^q-V_i(z_2))}{\theta ({\cal A}(P) +z_1U^p + z_2V^q + \zeta)}  + \cfrac{\sum\theta_i({\cal A}(P) +z_1U^p + z_2V^q + \zeta)V_i(z_2)}{\theta ({\cal A}(P) +z_1U^p + z_2V^q + \zeta)}\bigg) +
$$
$$
+O(\delta^2)+ O(\delta\varepsilon)\bigg)\bigg\vert_{P = p} =
\cfrac{\partial}{\partial k_1^{-1}} \ln \bigg(1 + \delta\bigg(\cfrac{\sum\theta_i({\cal A}(P)+ z_1U^p + z_2V^q + \zeta)V_i^q}{\theta ({\cal A}(P)+ z_1U^p + z_2V^q + \zeta)}\bigg)+O(\delta^2)+ O(\delta\varepsilon)\bigg)\bigg\vert_{P = p}=
$$
$$
=\cfrac{\partial}{\partial k_1^{-1}} \ln \bigg(1 + \delta\partial_{z_2}\ln\theta({\cal A}(P)+ z_1U^p + z_2V^q + \zeta)+O(\delta^2)+ O(\delta\varepsilon)\bigg)\bigg\vert_{P = p}=
$$
$$
=\cfrac{\delta\partial_{k_1^{-1}} \partial_{z_2}\ln\theta ({\cal A}(p)-U^pk_1^{-1}+O(k_1^{-2})+ z_1U^p + z_2V^q + \zeta)+O(\delta^2)+O(\delta\varepsilon)}{1 + \delta\partial_{z_2}\ln\theta({\cal A}(P)+ z_1U^p + z_2V^q + \zeta)+O(\delta^2)+ O(\delta\varepsilon)}\bigg)\bigg\vert_{P = p}=
$$
$$
= -\delta\partial_{z_1}\partial_{z_2}\ln\theta({\cal A}(p)+ z_1U^p + z_2V^q + \zeta)+O(\delta^2)+ O(\delta\varepsilon).
$$
Next, we have
$$
v(z_1,z_2)  = -\frac{1}{\delta} b(z_1,z_2) - g_{z_2+\delta}(0) + \partial_{z_1}\partial_{z_2}\ln\theta({\cal A}(p)+ z_1U^p + z_2V^q + \zeta)+ O(\varepsilon)+O(\delta).
$$

Now let us consider the expansion of the operator $L_{\varepsilon,\delta}.$ For computational convenience, we represent the shift operators $T_{\varepsilon}$ and $T_{\delta}$ in the following form:
$$
T_{\varepsilon}=1 + \varepsilon\partial_{z_1} +\tilde{T}_{\varepsilon}, \qquad 
T_{\delta}=1 + \delta\partial_{z_2} +\tilde{T}_{\delta},
$$ 
where $\tilde{T}_{\varepsilon}=O(\varepsilon^2), \ \tilde{T}_{\delta} = O(\delta^2).$ Then
$$
L_{\varepsilon,\delta} = \frac{1}{\varepsilon\delta}(1 + \delta\partial_{z_2} +\tilde{T}_{\delta})T_{\varepsilon} -\frac{1}{\delta} \frac{T_{\varepsilon}}{\varepsilon} + \frac{b(z_1,z_2)}{\delta}(1 + \delta\partial_{z_2} +\tilde{T}_{\delta}) -\frac{b(z_1,z_2)}{\delta} - g_{z_2+\delta}(0) + 
$$
$$
+\partial_{z_1}\partial_{z_2}\ln\theta({\cal A}(p)+ z_1U^p + z_2V^q + \zeta)+ O(\varepsilon) +O(\delta) = \frac{1}{\varepsilon}(\partial_{z_2} +\frac{\tilde{T}_{\delta}}{\delta})T_{\varepsilon} + b(z_1,z_2)(\partial_{z_2} +\frac{\tilde{T}_{\delta}}{\delta})-
$$
$$
-g_{z_2+\delta}(0) + \partial_{z_1}\partial_{z_2}\ln\theta({\cal A}(p)+ z_1U^p + z_2V^q + \zeta)+ O(\varepsilon) +O(\delta) = \frac{1}{\varepsilon}(\partial_{z_2} +\frac{\tilde{T}_{\delta}}{\delta})(1 + \varepsilon\partial_{z_1} +\tilde{T}_{\varepsilon}) +
$$
$$
- \big(\cfrac{1}{\varepsilon} + \partial_{z_1}\ln \bigg(\cfrac{\theta({\cal A}(q) + z_1 U^p + z_2 V^q + \zeta)}{\theta ({\cal A}(p) + z_1 U^p + z_2 V^q + \zeta)}\bigg) -s+ O(\varepsilon)+ O(\delta)\big)(\partial_{z_2} +\frac{\tilde{T}_{\delta}}{\delta})-g_{z_2+\delta}(0) +
$$
$$
+ \partial_{z_1}\partial_{z_2}\ln\theta({\cal A}(p)+ z_1U^p + z_2V^q + \zeta)+ O(\varepsilon) +O(\delta) = \frac{1}{\varepsilon}(\partial_{z_2} +\frac{\tilde{T}_{\delta}}{\delta})(\varepsilon\partial_{z_1} +\tilde{T}_{\varepsilon}) -
$$
$$
- \big(\partial_{z_1}\ln \bigg(\cfrac{\theta({\cal A}(q) + z_1 U^p + z_2 V^q + \zeta)}{\theta ({\cal A}(p) + z_1 U^p + z_2 V^q + \zeta)}\bigg) -s+ O(\varepsilon)+ O(\delta)\big)(\partial_{z_2} +\frac{\tilde{T}_{\delta}}{\delta})-g_{z_2+\delta}(0) +
$$
$$
+ \partial_{z_1}\partial_{z_2}\ln\theta({\cal A}(p)+ z_1U^p + z_2V^q + \zeta)+ O(\varepsilon) +O(\delta) =\partial_{z_1}\partial_{z_2}+ 
$$
$$
-\Big(\partial_{z_1}\ln \bigg(\cfrac{\theta({\cal A}(q) + z_1 U^p + z_2 V^q + \zeta)}{\theta ({\cal A}(p) + z_1 U^p + z_2 V^q + \zeta)}\bigg) -s\Big)\partial_{z_2}  +
\partial_{z_1}\partial_{z_2}\ln\theta({\cal A}(p)+ z_1U^p + z_2V^q + \zeta)+const+O(\delta)+ O(\varepsilon).
$$

Here $s$ and $g_{z_2+\delta}(0)$ are some constants. Then, for $z_1 = z$, $z_2 = \bar{z}$ we obtain $L_{\varepsilon,\delta} = H + \text{const} + O(\varepsilon) + O(\delta).$
Theorem 3 is proved.

\noindent{\bf Example 2}~\cite{LM}. For $g=1$, the coefficients of the operator $L_{\varepsilon, \delta}$ can be expressed in terms of Weierstrass elliptic functions:
$$
	a(z,\bar{z}) = -\cfrac{1}{\delta}, 
$$
$$
	b(z,\bar{z}) = \zeta (p - q) + \zeta (q - \gamma_1(z) - \gamma_2(\bar{z}+ \delta)) - \zeta (\gamma_1(z + \varepsilon) - \gamma_1(z) ) - \zeta (p - \gamma_1(z + \varepsilon) - \gamma_2 (\bar{z} + \delta)) , 
$$
$$
	v(z,\bar{z}) = \lambda(z + \varepsilon, \bar{z}) ( \wp (p - q) - \wp (p - \gamma_1(z + \varepsilon) -\gamma_2(\bar{z}) ) ) - \cfrac{b(z,\bar{z})}{\delta} . 
$$
Let $\gamma_1(z)$ and $\gamma_2(\bar{z})$ have the expansions 
$$
	\gamma_1 (z) = z + c \gamma  + \alpha_2 (z) \varepsilon^2 + \ldots,  \quad
	\gamma_2(\bar{z}) = \bar{z} + (1 - c) \gamma + \beta_2(\bar{z}) \delta^2 + \ldots.
$$
Then  
$$
	\lim\limits_{\varepsilon, \delta \to 0} L_{\varepsilon, \delta} = H,
$$
where $H$ is the finite-gap Schr\"odinger operator from Example 1.

P.A. Leonchik,

Faculty of Mathematics, Technion - Israel Institute of Technology, Technion City, Haifa, 320000, Israel

Department of Mathematics and Computer Science, Guangdong Technion - Israel Institute of Technology, 241 DaXue Road, Shantou, Guangdong, 515063, China

e-mail: leonchik.2002@mail.ru

${\ }$

G.S. Mauelshova,

S.L. Sobolev Institute of Mathematics, Siberian Branch of Russian Academy of Sciences, 4 Acad. Koptyug Ave., Novosibirsk, 630090, Russia

Novosibirsk State University, 1 Pirogova St., Novosibirsk, 630090, Russia

e-mail: mauleshova@math.nsc.ru

${\ }$

A.E. Mironov,

S.L. Sobolev Institute of Mathematics, Siberian Branch of Russian Academy of Sciences, 4 Acad. Koptyug Ave., Novosibirsk, 630090, Russia

Novosibirsk State University, 1 Pirogova St., Novosibirsk, 630090, Russia

e-mail: mironov@math.nsc.ru


\begin{thebibliography}{1}
\bibitem{MM}
G.~S.~Mauelshova, A.~E.~Mironov, {\it One-dimensional finite-gap Schr\"odinger operators as a limit of commuting difference operators}, Dokl. Math., {\bf 512} (2023), 81--84.

\bibitem{KN}
I.~M.~Krichever, S.~P.~Novikov, {\it Two-dimensionalized Toda lattice, commuting difference operators, and holomorphic bundles}, Russian Math. Surveys, {\bf 58}:3 (2003), 51--88.

\bibitem{DKN}
B.~A.~Dubrovin, I.~M.~Krichever, S.~P.~Novikov, {\it The Schr\"odinger equation in a periodic field and Riemann surfaces}, Soviet Math. Dokl., {\bf 229}:1 (1976), 15--18.

\bibitem{NV}
A.~P.~Veselov, S.~P.~Novikov, {\it Finite-gap two-dimensional Schr\"odinger operators. Potential operators}, Soviet Math. Dokl., {\bf 279}:4 (1984), 784--788.

\bibitem{GMN1}
Grinevich, P.G., Mironov, A.E. Novikov, S.P. Zero level of a purely magnetic two-dimensional nonrelativistic Pauli operator for SPIN-1/2 particles. Theor Math Phys 164, 1110–1127 (2010).

\bibitem{GMN2}
P. G. Grinevich, A. E. Mironov, S. P. Novikov, “On the non-relativistic two-dimensional purely magnetic supersymmetric Pauli operator”, Russian Math. Surveys, 70:2 (2015), 299–329

\bibitem{LM}
P.~A.~Leonchik, A.~E.~Mironov, {\it Two-dimensional discrete operators and rational functions on algebraic curves}, Sao Paulo Journal of Mathematical Sciences, {\bf 18} (2024), 855--865.

\bibitem{K}
I.~M.~Krichever, {\it Two-dimensional periodic difference operators and algebraic geometry}, Soviet Math. Dokl., {\bf 285}:1 (1985), 31--36.

\bibitem{DGNS}
A.~Doliwa, P.~G.~Grinevich, M.~Nieszporski, P.~M.~Santini, {\it Integrable lattices and their sub-lattices: from the discrete Moutard (discrete Cauchy-Riemann) 4-point equation to the self-adjoint 5-point scheme}, Journal of Mathematical Physics, {\bf 48}:1 (2007), 013513, 28 pp.

\bibitem{N}
Novikov, S.P. The periodic problem for the Korteweg—de vries equation. Funct Anal Its Appl 8, 236–246 (1974). 

\bibitem{IM}
Its, A.R., Matveev, V.B. Schrödinger operators with finite-gap spectrum and N-soliton solutions of the Korteweg-de Vries equation. Theor Math Phys 23, 343–355 (1975).

\bibitem{DMN}
B. A. Dubrovin, V. B. Matveev, S. P. Novikov, “Non-linear equations of Korteweg–de Vries type, finite-zone linear operators, and Abelian varieties”, Russian Math. Surveys, 31:1 (1976), 59–146

\bibitem{Dub1}
B.~A.~Dubrovin, {\it Riemann surfaces and nonlinear equations}, Izhevsk: Regular and Chaotic Dynamics, 2001, 152 pp.

\end{thebibliography}
\end{document}